\documentclass[11pt,preprint]{aastex}

\newcommand{\AFGL}{AFGL~2591}
\newcommand{\kms}{km s$^{-1}$}
\newcommand{\HII}{\hbox{H~{\sc ii}}}
\usepackage{lscape}

\begin{document}

\title{Clustered star formation and outflows in AFGL~2591}

\author{A. Sanna\altaffilmark{1}, M. J. Reid\altaffilmark{2}, C. Carrasco-Gonz\'alez\altaffilmark{1},
K. M. Menten\altaffilmark{1}, A. Brunthaler\altaffilmark{1}, L. Moscadelli\altaffilmark{3}, K.L.J. Rygl\altaffilmark{4}}

\email{asanna@mpifr-bonn.mpg.de}
\altaffiltext{1}{Max-Planck-Institut f\"{u}r Radioastronomie, Auf dem H\"{u}gel 69, 53121 Bonn, Germany}
\altaffiltext{2}{Harvard-Smithsonian Center for Astrophysics, 60 Garden Street, Cambridge, MA 02138, USA}
\altaffiltext{3}{INAF, Osservatorio Astrofisico di Arcetri, Largo E. Fermi 5, 50125 Firenze, Italy}
\altaffiltext{4}{IFSI-INAF, Istituto di Fisica dello Spazio Interplanetario, Via del Fosso del Cavaliere 100, 00133 Roma, Italy}
\begin{abstract}

We report on a detailed study of the water maser kinematics and radio continuum emission toward
the most massive and young object in the star-forming region \objectname{AFGL~2591}.
Our analysis shows at least two spatial scales of multiple star formation, one projected across 0.1~pc on
the sky and another one at about $\rm 2000~AU$ from a ZAMS star of about $\rm 38~M_{\odot}$. This young stellar
object drives a powerful jet- and wind-driven outflow system with the water masers associated to
the outflow walls, previously detected as a limb-brightened cavity in the NIR band. At about 1300~AU to the
north of this object a younger protostar drives two bow shocks, outlined by arc-like water maser emission,
at 200~AU either side of the source. We have traced the velocity profile of the gas that expands along
these arc-like maser structures and compared it with the jet-driven outflow model. This analysis suggests
that the ambient medium around the northern protostar is swept up by a jet-driven shock ($\rm >66~km~s^{-1}$)
and perhaps a lower-velocity ($\rm \sim 10~km~s^{-1}$) wind with an opening angle of about $20\degr$ from the jet axis.

\end{abstract}


\section{Introduction}

The study of high-mass ($>8~\rm M_{\odot}$) star formation still faces fundamental questions such as (e.g.,
\citealt{Zinnecker2007,Beuther2007}) what is
the origin of the mass reservoir for the protostar (whether the protostar can gather mass only
from its own cocoon or competes with the nearby protostellar companions to accrete mass from the whole molecular clump)
and what is the accretion mechanism (whether the mass is conveyed to the protostar through a stable accreting disk
or via episodic accreting flows while the protostar crosses the denser portions of the molecular clump).
To examine these questions, observations on linear scales matching the sizes of the molecular clumps (0.1--1~pc)
need to be combined with a closer view of the protostellar surroundings (the hot molecular cores, HMCs,
with sizes of a few thousands of AU) to establish the presence of multiple sources, their kinematical
structures, and a possible overall interaction.

Focusing on a single protostellar core, Very Long Baseline Interferometry (VLBI) proper motion studies of maser
spots in conjunction with sub-arcsecond imaging of the free-free emission of very young \HII\ regions and thermal jets
have proven a powerful tool to investigate the kinematics and the structure of the gas close to the protostar(s).
For instance, our VLBI multi-epoch observations toward the massive young stellar object (YSO)
G23.01-0.41 have shown three different kinematical structures traced by the H$_2$O, OH, and CH$_3$OH
masing molecules within 2000~AU from the central YSO. While the water masers trace a bipolar flow associated
with a thermal jet, the hydroxyl masers belong to an expanding layer in front of the central source.
Furthermore, the methanol masers appear to undergo both rotation and expansion inside a toroidal structure
traced in NH$_3$ lines and extended over about 0.1~pc \citep{Sanna2010b}.

The present paper focuses on the water masers and radio continuum toward the high-mass star-forming region \AFGL.
The distance to this region has been substantially underestimated in the past and we have recently measured
an accurate value of 3.3~kpc with the trigonometric parallax of its 22.2~GHz H$_2$O masers (Rygl et al. 2011, submitted).
Based on the data of \citet{Minh2008} and \citet{Lada1984}, we find
a revised value for the large-scale ($<10$~pc) clump mass of about $\rm 2\times10^4~M_{\odot}$
and a bolometric luminosity inferred from IR data of $\rm 2\times10^5~L_{\odot}$. From
observations of the $^{13}$CO rovibrational lines, the systemic velocity of the region (V$_{\rm sys}$) with respect
to the local standard of rest (LSR) is $-5.7$~\kms\ \citep{vanderTak1999}.

\AFGL\ shows energetic star formation activity with hot core emission detected in several molecular lines (e.g.,
\citealt{vanderTak1999,Doty2002,Benz2007,vanderWiel2011}). A number of compact radio continuum sources were resolved in the region
\citep{Campbell1984,Trinidad2003,vanderTak2005} and associated with distinct clusters of water masers \citep{Tofani1995,Trinidad2003},
which establishes the occurrence of multiple star formation. \AFGL\ exhibits a powerful CO outflow, which is extended more than
$5'\times 5'$ ($\rm \sim 5~pc \times 5~pc$), with the blue lobe toward the west and the red lobe to the northeast (e.g.,
\citealt{Lada1984,Hasegawa1995}).
This outflow activity is also associated with several Herbig-Haro objects toward the west \citep{Poetzel1992} and H$_2$ bow shocks
\citep{Tamura1992}, marking the interaction between outflowing material and the surrounding molecular envelope.
There is evidence for an outflow cavity that enhances the chemistry along the outflow walls far from the dust peak
(e.g., \citealt{Bruderer2009}). This outflow cavity is observed as NIR loops, associated with the blueshifted
outflow lobe, at the high resolution of 170~mas by \citet[][cf. also \citealt{Tamura1991,Minchin1991}]{Preibisch2003}.
The apex of these NIR loops is centered on the most compact radio continuum source in the region called
VLA--3 (after \citealt{Trinidad2003}).

The observations reported here focus on the water maser emission within $2''$ from VLA--3.
In Sect.~2, we describe our Very Long Baseline Array (VLBA) observations of the 22.2~GHz H$_2$O masers together with the
archival Very Large Array (VLA) A-configuration observations of the radio continuum emission at 1.3 and 0.7~cm. In Sect.~3,
we illustrate the spatial morphology and kinematics of the maser emission and present results from archival VLA observations,
constraining the properties of the radio continuum associated with the masers. Section~4 discusses the properties of each cluster
of water maser emission detected in the region and draws an overall view for the purposes of star formation. The main conclusions
are summarized in Sect.~5.

\section{Observations and Data Analysis}

\subsection{VLBA observations: 22.2~GHz H$_2$O masers}

We conducted VLBA\footnote{The VLBA is operated by the National Radio Astronomy Observatory (NRAO). The NRAO is a facility of the National Science
Foundation operated under cooperative agreement by Associated Universities, Inc.} observations in the K band to study the $6_{16}-5_{23}$ H$_2$O
maser emission (rest frequency 22.235079 GHz) toward \objectname{AFGL~2591}.
%(tracking center: $\rm R.A.(J2000) = 20^h29^m24\fs8220$ and $\rm Dec.(J2000) = 40\degr11'19\farcs723$)
The VLBA observations were scheduled under program BM272H at four epochs: 2008 November 10, 2009 May 6, 13, and November 13.
While these observations were optimized to measure the annual parallax and Galactic proper motion of the source (Rygl et al. 2011,
submitted), in this paper we focus on the internal kinematics of the water maser emission.

We performed phase-referencing observations by fast switching between the maser target and the ICRF calibrator J2007+4029. That allowed us to
determine the water masers absolute position within an accuracy of $\pm 1$~mas. Two fringe finders (3C345 and 3C454.3) were observed for
bandpass, single-band delay, and instrumental phase-offset calibration. Further details about these observations as well as the distance
measurement to AFGL~2591 can be found in Rygl et al. (2011, submitted).
We employed four adjacent IFs of 8~MHz bandwidth in dual circular polarization, each one split into 256 spectral channels. This receiver setup
provided both a large-enough bandwidth to increase the SNR of the weak continuum calibrators and a 0.42~\kms\ channel width to adequately
sample the maser lines. The third IF centered on an LSR velocity (V$_{\rm LSR}$) of -5.0~\kms\ covered the previous detections of water maser
emission in the region \citep{Trinidad2003}. The data were processed with the VLBA correlation facility in Socorro (New Mexico) using an averaging
time of about 0.9~s which limited the instantaneous field of view of the interferometer to about $2''$ (i.e. without significant amplitude losses).
Data were reduced with the NRAO Astronomical Image Processing System (AIPS) following the procedure described in \citet{Reid2009a}.

\begin{figure*}
\centering
\includegraphics[angle= 0, scale= 0.4]{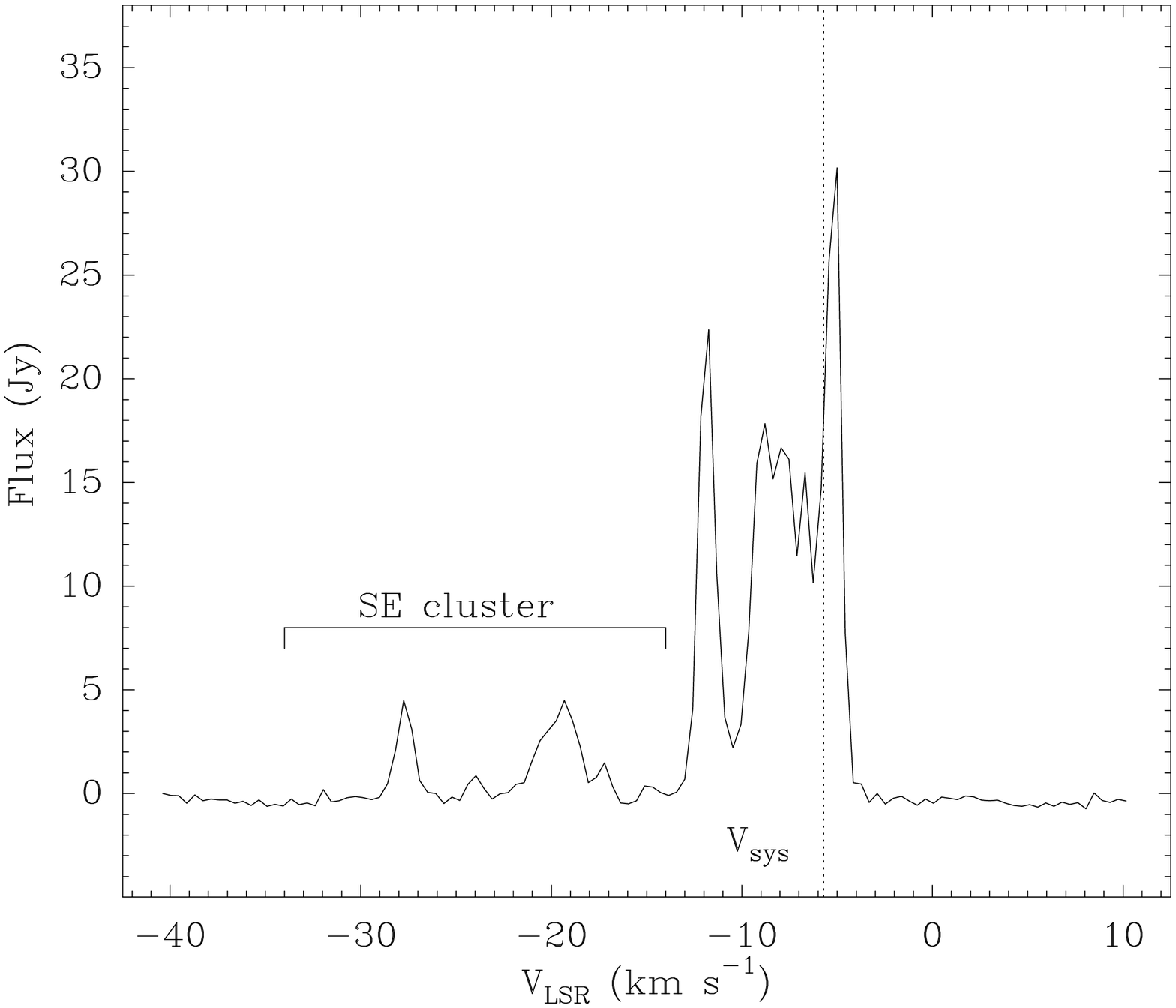}
\caption{Total-power (Stokes~I) spectrum of the 22~GHz H$_2$O masers toward AFGL~2591 from the second (middle) epoch on 2009 May 6. This profile was
produced by averaging the total-power spectra of all VLBA antennas, after weighting each spectrum with the antenna system temperature. The dotted
line crossing the spectrum (at $\rm -5.7~km~s^{-1}$) represents the systemic velocity (V$_{sys}$) of the molecular cloud hosting the star-forming
region. The LSR velocity range from the SE water maser cluster is shown. \label{spectra}}
\end{figure*}

The natural CLEAN beam was an elliptical Gaussian with an FWHM size of $\rm 0.80~mas \times 0.40~mas$ at a P.A. of -14.9\degr\ (east of north),
with small variations from epoch to epoch. In each observing epoch, with an on-source integration time of about 2~h, the effective rms
noise level on the channel maps was about 0.01~Jy~beam$^{-1}$. The total-power spectrum of the 22.2~GHz masers toward \AFGL\ is shown in
Figure~\ref{spectra}.

\subsection{VLA archival data}

We retrieved VLA A--configuration observations toward \AFGL\
at 1.3 and 0.7~cm from the VLA\footnote{The VLA is operated by the NRAO.} archive for the purposes of a detailed comparison with the high-resolution VLBA maser data. These observations were scheduled on 1999 June 29
\citep{Trinidad2003} and 2002 March 23 \citep{vanderTak2005}, respectively. We recalibrated these data with the NRAO AIPS software package using
standard procedures and cross-calibration between the strong H$_2$O maser lines and the continuum at 1.3~cm (e.g., \citealt{Reid1990}).
For the 1.3~cm data, we cleaned the VLA maps with a circular restoring beam, equal to the natural-beam minor axis for a ``ROBUST 0''
weighting (Table~\ref{cont}).

\section{Results}

\subsection{22.2~GHz H$_2$O masers}

We imaged a range of LSR velocities between -40 to +25~\kms\ with a field-of-view of
$ 1\farcs4 \times 1\farcs4 $ about the VLA--3 radio source.
These limits include the emission previously reported by \citet[][their Table~3 and~4]{Trinidad2003} toward VLA--3.
In the following, the term \emph{feature} refers to spots spatially overlapping in contiguous velocity channels, that
we treat as an individual masing cloud (see \citealt{Sanna2010a} for a detailed discussion).

We identified 80 distinct water maser features distributed within an area of about $0\farcs77 \times 0\farcs48$
($\rm \sim 2500~AU \times 1500~AU$). They are mainly grouped in five clusters labeled in Table~\ref{distrib} north-west (NW),
middle-west (MW), north-east (NE), middle-east (ME), and south-east (SE) in Figure~\ref{puzzle}a.
The properties of individual features are presented in Table~\ref{distrib} and show peak brightness ranging from 22~Jy~beam$^{-1}$ to
a detection limit of 0.05~Jy~beam$^{-1}$ ($5\sigma$).  Water maser features redshifted with respect to the V$_{sys}$ (of $-5.7$~\kms)
belong to the MW cluster, whereas the blueshifted emission is clustered toward the SE. The full spread in LSR
velocities ranges from $-0.4$~\kms\ for the most redshifted feature~(\#~13) to $-34$~\kms\ for the most blueshifted one (\#~77).
The angular distribution and full-space motions of the water masers (i.e. line-of-sight velocities plus proper motions) are plotted
in Figure~\ref{puzzle}a. With a time baseline of 1~yr, we measured relative proper motions of individual features
with an uncertainty less than 3~\kms.
Positions and velocities in Table~\ref{distrib} are relative to the compact feature~\#~4
which was stable in the spectral domain over the 1~yr observations. The magnitude of relative proper motions ranges from 5.0~\kms\
for feature~\#~17 to 56~\kms\ for feature~\#~54.
Relying on the symmetry of the maser spatial distribution, we have corrected the proper motions by the average velocity of
all features of V$_{\rm x}=8.8\pm0.4$ and V$_{\rm y}=9.0\pm0.4$, to approximate the actual motions with respect to the center
of expansion.

In Figure~\ref{puzzle}b, \ref{puzzle}c, and \ref{puzzle}f, we zoom in on the three water maser clusters
labeled MW, NE, and SE, showing a symmetric expansion away from their centroids.
The NE and SE clusters are particularly interesting due to their full kinematics and are discussed in detail
in Section~\ref{NEcluster} and~\ref{SEcluster}, respectively.

In Figure~\ref{puzzle}d--e, we overlay the radio continuum emission described in Section~\ref{radiocontinuum} with
the water maser emission from the second epoch observations. This maser emission was summed over the full velocity
range of the SE cluster (from $V_{\rm LSR}$ of $-34$ to $-14$~km~s$^{-1}$).
Its projected shape along the plane of the sky delineates a straight ``V'' (hereafter the H$_2$O V--shape) with a
full opening angle of $110\degr \pm 10\degr$ and a P.A. of 256\degr.
The variation of LSR velocities along this V--shape is quite regular, ranging from a maximum of $-13$~\kms\ (feature \#~78)
at the northern side of the V--shape to a minimum of about $-33$~\kms\ (feature \#~77) toward the southern side.

\subsection{Radio continuum emission}\label{radiocontinuum}

Our measurements of the 1.3 and 0.7~cm continuum emissions are summarized in Table~\ref{cont} and plotted in
Figure~\ref{puzzle}d--e. Published VLA C-configuration observations at 0.7~cm show that significant emission is resolved out with
the VLA A-configuration, missing more than half of the flux measured in the compact configuration \citep{vanderTak2005}.
The relative position of the 1.3~cm continuum with respect to the reference maser feature in Table~\ref{distrib}
is obtained by matching the overall water maser distribution between the VLBA and VLA--A maps. We note that the NS
spread of the maser emission overlaid on the radio continuum ($\rm \sim 50~mas$) have remained almost stationary over
about ten years (cf. \citealt{Trinidad2003}).

The peak position of the 0.7~cm continuum is assumed to be aligned with
the 1.3~cm peak. The three-lobed geometry of the 7~mm emission follows the 1.3~cm elongation toward the east and
it branches off toward the west matching the H$_2$O V--shape (Table~\ref{cont}). While this matching is particularly
striking, we note that more sensitive, Q--band, observations are mandatory to confirm the spatial structure
of the 0.7~cm emission. Because of the stability of the water maser distribution, but also the similarities between
the morphology of the water maser emission and the 0.7~cm continuum (Figure~\ref{puzzle}e), we estimate a
relative position accuracy between the radio continuum and water masers of $\pm 0\farcs01$.

Previous observations spanning a broad-band spectrum between 6~cm to 1.3~mm show that the spectral index ($S\propto\nu^\gamma$) for wavelengths longer than 7~mm is
dominated by free-free emission ($\gamma \sim 0.9$), whereas at shorter wavelengths the flux density is mostly due to dust emission
($\gamma > 2$; \citealt{vanderTak2005} and reference therein).

\section{Discussion}

\subsection{Outflow emission in the NE H$_2$O cluster}\label{NEcluster}

In this section, we discuss two basic characteristics observed in the NE water maser cluster (Figure~\ref{puzzle}b).
Firstly, we detected two arc-like filaments of water maser emission elongated in the NE--SW direction with no emission
in between them. Secondly, these arc-like filaments expand from each other with proper motions increasingly
aligned with the NE--SW direction as one approaches this direction.
Bipolar distributions of water masers with expanding proper motions have been reported in a large number of star-forming
regions (e.g., \citealt{Gwinn1992,Sanna2010b,Moscadelli2011}). These detections were interpreted as footprints of the
interaction between collimated jets and/or wide-angle winds from YSOs with the surrounding molecular envelope.
Following this evidence, we want to show that the overall kinematics of the NE water maser cluster can be explained by
assuming that a single object, in between the maser emission, emits a high-velocity, bipolar jet along the NE--SW direction
(Figure~\ref{puzzle}b and \ref{pv}a).
At first, we will compare the velocity profile of the water masers along the arc-like structures with the model of jet-driven
outflows presented in \citet{Ostriker2001} and \citet{Lee2001}. The jet model predicts that the interaction between
a bow shock and the ambient material gives an upper and lower limit to the velocity of the processed material. This
processed material is a mixture of material that was already part of the bow shock and material incorporated from the
ambient medium as the shock front expands. The shock front delineates an arc-like structure (as do the masers) and the velocity component
along the jet axis decreases roughly as $r^{-2/3}$, where $r$ is the offset distance from the head of the jet (e.g., Figure~2 in
\citealt{Ostriker2001}).

The similarities of the spatial elongation and velocities of the water masers would be naturally explained if the NE--SW direction
is associated with an outflow in this direction. In Figure~\ref{pv}a, we define an elliptical symmetry that matches to first order
both the width and the NE-SW elongation of the water maser distribution with a simple central symmetry (major axis and eccentricity on
top of Figure~\ref{pv}a). This elliptical pattern allows us to define a new reference
frame with origin at the center of the ellipse ($\Delta \rm x = 625$~mas and $\Delta \rm y = 260$~mas w.r.t. the reference feature in
Table~\ref{distrib}) and \emph{x}, \emph{y} axes along the minor (R) and major (z) axes of the ellipse, respectively.
This reference frame is tilted by $22\degr$ to the east of the north direction. Due to the small dispersion of LSR velocities (1.5~\kms)
centered at about $-7.9$~\kms, we also assume that all the maser features lie along the same plane. We want to study the velocity profile
of the shock fronts traced by the water maser proper motions with respect to the new \emph{z--R} reference frame. In the following,
we use the same terminology of \citet{Ostriker2001} and \citet{Lee2001} in order to facilitate the comparison with their jet model.

In Figure~\ref{pv}b, we report the behavior of the velocity components (V$_{\rm z}$) of the water masers along the major axis of the
ellipse (z). The idea that the z-direction is a preferred outflow direction is supported by the velocity profile in Figure~\ref{pv}b,
which resembles that expected for a ballistic bow-shock produced by a jet along the z-direction (cf. Figure~4 of \citealt{Lee2001}).
For such a bow-shock model velocities are constrained in Figure~\ref{pv}b by two solid lines and the expected gas velocity should lie between
them. These limits are derived from equations (22), (18), and (20) of \citet{Ostriker2001} and define the bow shock profile $z=f(R)$,
the mean velocity of the mixing of jet and ambient material $v'_{z}$ (solid line) in a shell, and the velocity of newly shocked ambient
material $u'_{z}$ (dashed steeper line), respectively. These equations are parameterized by the jet radius $\rm R_j$, the velocity of the
bow shock $\rm v_s$, and the ejection velocity  $\rm \beta c_s$ of the material processed by the working surface, expressed as a
multiple of the isothermal sound speed \citep{Ostriker2001}.  For the radius of the high-velocity ``core'' of the jet ($\rm R_j$), we set the
minimum projected distance to the R-axis measured with the masers of 1.8~mas ($\approx 6$~AU). This value implies a width-to-length ratio
of the jet from the center of the ellipse of 0.05, consistent with that found in the literature for H$_2$ jets at typical scales
of $10^3-10^4$~AU (e.g., HH~212 in \citealt{Zinnecker1998}). We stress that $\rm R_j$ is the only constraint we put on the model, whereas
we consider the best value of the ratio $\rm \beta c_s/v_s = 0.5$ and $\rm \beta = 4.1$ inferred from the simulations by \citet{Lee2001}.
By setting the tip of the bow shock at the maximum maser projection along z (the ``0'' in \citealt{Ostriker2001}), we find
a good agreement of the measured maser velocities (black squares in Figure~\ref{pv}b) with the jet-driven outflow model. We note that the
maser velocities are closer to the ambient material velocity  (dashed line) in agreement with simulations by \citet[their Figure~4]{Lee2001}.
This behavior may reflect an incomplete mixing of material already in the bow shock shell and material added from
the ambient medium. The $\rm \beta c_s/v_s$ ratio implies a bow shock speed of about 66~\kms\ for a sound speed of 8~\kms\
at the cooling cutoff of $10^4$~K (see \citealt{Ostriker2001}). On the contrary, in the alternative scenario of a wide-angle wind
the V$_{\rm z}$ motions would increase linearly with z, that is not observed (e.g., \citealt{Lee2001} their Figure~9).
This suggests that the maser cloudlets are dragged along the z-direction by a jet-driven shock component.

On the other hand, the jet-driven outflow model predicts an upper limit to the transverse velocity component V$_{\rm R}$
with respect to the jet direction of the order of the sound speed ($\sim 8$~\kms; see Figure~3 in
\citealt{Ostriker2001} and  Figure~4 in \citealt{Lee2001}). This limit is intimately related to the assumption that the
transverse radial forces are dominated by thermal pressure gradients that act at the head of the jet.
In Figure~\ref{pv}c, we measure the greatest deviations from the jet model far from the head of the jet, at an inclination
of about $20\degr$ from the jet axis. This behavior may suggest that, away from the head of the jet, thermal pressure gradients
no longer dominate the expansion. One possibility would be that a low-velocity ($< 10$~\kms) wind with
a wide opening angle (ca. $20\degr$) acts with the jet simultaneously. In the wind-driven outflow model, the expansion of
the outflow away from the central axis is driven by the ram pressure of a wide-angle wind \citep{Shu1991}:
the V$_{\rm R}$ velocity is maximum far from the tip of the outflow shell and goes to zero at its tip (Figure~9 in \citealt{Lee2001}).
The simultaneous presence of the two types of outflow mechanisms would agree with the hypothesis that any outflow contains
a wide-angle wind at some level, in order to explain typical width-to-length ratio of about $1:10$, much higher than in a ``pure jet
model'' (e.g., \citealt{Arce2007}).

Finally, if the single jet-driven model is correct, we can also get an estimate of the dynamical age of the jet-event exciting the water masers. The
velocity of the jet ($\rm v_j$) is related to the bow shock speed ($\rm v_s$) by the ratio of the jet-to-ambient density, where $\rm v_s$
approximates $\rm v_j$ if the jet density exceeds the ambient density (see \citealt{Ostriker2001}). By considering the lower limit
$\rm v_j \thickapprox  v_s$, the inferred value of $\rm v_s = 66$~\kms, and the value of the higher maser projection along z,
we obtain a dynamical age of $\rm t=z_{maj}/v_j\backsimeq 14$~yr.
On the one hand, this dynamical age is consistent with other estimates based on the expanding motions
of masers around forming YSOs (e.g., \citealt{Torrelles2003}, \citealt{Moscadelli2007}). On the other hand, given how unlikely
it would be to observe such a short-lived event, it strongly suggests that we are tracing a recurrent phenomenon of ejection of matter from an YSO.
That would favor a pulsed-jet paradigm against a steady-jet one as supported, for instance, by the recurrent knots and bow shocks observed in
the prototypical Herbig-Haro object 212 \citep{Zinnecker1998}.
This interpretation naturally explains also the eastern water masers in Figure~\ref{pv}a (features \# 38, 39 in Table~\ref{distrib})
and the masers at about 150~mas to the south of the NE cluster (the ME cluster in Table~\ref{distrib}). These masers would belong to
an older jet emission from the same object.

\subsection{Outflow emission in the SE H$_2$O cluster associated with VLA--3}\label{SEcluster}

In this section, we discuss three characteristics observed in the SE water maser cluster:
\textrm{I.} the projected V--shape of the water maser distribution on the plane
of the sky, ``pointing'' toward the peak of the radio continuum emission (Figure~\ref{puzzle}d--e);
\textrm{II.} the water maser proper motions that mainly expand along the V--shape (Figure~\ref{puzzle}a);
\textrm{III.} the line-of-sight velocity gradient observed through the V--shape, with velocities increasing to
the south (Figure~\ref{puzzle}f).
In the following, we relate these properties with the large-scale outflow and the nature of the
radio continuum emission and suggest that the H$_2$O masers trace the edges of the outflow cavity
associated with a massive YSO (see Figure~\ref{outflow}a).

The H$_2$O emission from the SE cluster  is blueshifted with respect to the overall velocity of the cloud
($\rm -5.7~km~s^{-1}$) and covers a range of line-of-sight velocities consistent with the small-scale, blueshifted wing
of the CO(3--2) outflow (between -10 and -40~\kms; \citealt{Hasegawa1995}).
The V--shape observed in the water maser emission opens toward the west, where a limb-brightened cavity
was previously outlined as NIR loops \citep{Tamura1991,Minchin1991,Preibisch2003}. These loops were explained as
wind-bubbles, which were created as the blueshifted (western) side of the outflow pushed against the environment and
swept up ambient material \citep{Preibisch2003}. We note that the wide opening angle of the NIR loops ($> 100\degr$)
is consistent with that of the V--shape ($110\degr$). Actually, if the loops outlined the outflow
cavity, one might expect the water masers along the outflow walls, where shocks are more likely.
The H$_2$O V--shape opens toward a position angle (P.A.=256\degr) in good agreement with the direction of the western side of
the outflow, which has been detected in a number of tracers: \textrm{i)} the NIR bright loops, with a $262\degr \leq \rm P.A. \leq 256\degr$,
surrounding the outflow cavity with a symmetry axis at a P.A. of 259\degr\ (e.g., \citealt{Preibisch2003}, see their Figure~3);
\textrm{ii)} the blueshifted CO lobe with a P.A. of about 260\degr\ (e.g., \citealt{Hasegawa1995}, see their
Figure~4); \textrm{iii)} the western, aligned, Herbig-Haro objects with a $258\degr \leq \rm P.A. \leq 261\degr$ \citep{Poetzel1992}
and the hot H$_2$ knots in the same direction \citep{Tamura1992}; \textrm{iv)} the elongated 1.3~cm continuum with
a P.A. of 268\degr\ and a spectral index consistent with an ionized wind (Figure~\ref{puzzle}d).
If water masers would belong to the blueshifted outflow lobe, and might possibly amplify the continuum background at 1.3~cm,
it would naturally explain the non-detection of redshifted maser emission, that would belong to the receding outflow lobe.
Also, since the outflow axis is close to the line-of-sight (within $30\degr-45\degr$; \citealt{Minchin1991,Hasegawa1995,vanderTak1999}),
this scenario agrees with the direction of the maser proper motions, that are mainly expanding parallel to the V--shape with a strong
component toward the observer (Figure~\ref{outflow}). The velocity gradient observed across the maser V--shape (Figure~\ref{puzzle}f)
might be reproduced with a slow precession of the outflow axis about the mean direction $\rm Z_{out}$ in Figure~\ref{outflow}
(as speculated at first by \citealt{Trinidad2003}).

The strong water masers and the radio continuum emission from VLA--3 (Figure~\ref{puzzle}d--e) suggest an association with
a massive YSO \citep{vanderTak2005}. If the free-free emission longward of 7~mm wavelength originates from an ionized wind, we
can estimate a lower limit to the mass-loss rate ($ \dot{M}_w$) of the wind and put a constraint to the the nature of the driving
source. We apply equation (4) of \citet{Rodriguez1983} and use the flux at 1.3~cm to estimate the mass-loss
rate in the assumption of a spherical, isothermal, completely photoionized, uniform wind:

{\footnotesize
\begin{eqnarray}\label{massloss}
\biggr(\frac{ \dot{M}_w}{\rm 10^{-5}~M_{\odot}~yr^{-1}}\biggr) & \approx &  1.7 \biggr(\frac{S_{\nu}}{\rm 1.6~mJy}\biggr)^{3/4}
\biggr(\frac{\nu}{\rm 22.2~GHz}\biggr)^{-0.45} \nonumber \\
& & \cdot \biggr(\frac{d}{\rm 3.3~kpc}\biggr)^{3/2}\biggr(\frac{v_w}{\rm 10^3~km~s^{-1}}\biggr),
\end{eqnarray}
}
where $ v_w$ is the velocity of the wind. We have considered the accurate value of the recently measured distance to AFGL~2591 and a wind
velocity of several $10^2$~\kms, reported in both the $\rm [S II]$ lines by \citet{Poetzel1992} and the IR $^{12}$CO broad wings by
\citet{vanderTak1999}. The assumption of a spherical wind is justified by the wide opening angle observed both with the maser emission
and the NIR loops ($> 100\degr$).
Note that, \citet{Trinidad2003} used the lower flux at 3.6~cm  to estimate the mass-loss rate, based on their model with a dusty disk
that would have affected the flux at 1.3~cm. Still, at the new measured distance of 3.3~kpc, their modeled flux would decrease
by a factor 10 and no longer be applicable, while the spectral index for wavelengths longer than 0.7~cm suggests an origin
in a ionized wind \citep{vanderTak2005}. The mass-loss rate from equation~(\ref{massloss}) implies a photoionizing ZAMS star of type
O9, with a rate of ionizing photons (N$_L$) of about $\rm 1.2\times10^{48}~s^{-1}$ \citep{Rodriguez1983}. We compare this value with
the estimates of mass and Lyman continuum flux derived by assuming a single star dominating the  IR luminosity of the region.
Following \citet{vanderTak2005}, we used the theoretical HR diagram by \citet{Maeder1989} to estimate the mass and effective temperature
of a star with an IR luminosity of $\rm 2\times10^5~L_{\odot}$; then, we used the stellar atmosphere models by \citet{Schaerer1997}
to convert effective temperatures to Lyman fluxes. These new values of M and N$_L$ are about $\rm 38~M_{\odot}$
and $\rm 2\times10^{49}~s^{-1}$, respectively. The estimates of Lyman flux obtained independently from the radio continuum emission
and the IR luminosity of the region agree within an order of magnitude. Since the former one is a lower limit derived close to the source
whereas the later one gives an upper limit to the number of Lyman photons emitted over a larger field of view, this calculation provides
evidence that a single massive object dominates the IR energetics of the region (according to the high-resolution NIR image of
\citealt{Preibisch2003}).

\subsection{Clustered star formation around VLA--3}

We finally collect the information from the water masers and the radio continuum emission
about the ongoing star formation in the region. Taking into account a possible bias of detecting
only a small number of (bright) sources, there is evidence of \emph{at least} two clustering scales
for star formation in the region. One on a linear scale of about 0.1~pc on the sky, from continuum
sources VLA--1, VLA--2, and VLA--3 detected at first by \citet[][named by
\citealt{Trinidad2003}]{Campbell1984}. The other scale is an order of magnitude smaller and comes from our observation
of at least three centers of expansion that pinpoint two other centers of star formation (the NE and MW cluster),
separated by about $\rm 2000~AU$ on the sky from the VLA--3 radio source. Note that, the NW maser cluster might be
either associated with the MW cluster or trace a further young object.
Our analysis in Section~\ref{SEcluster} shows that VLA--3 is the most massive ($\rm \lesssim 38~M_{\odot}$) and young
object in the field dominating the IR luminosity. Several lines of arguments from the spatial scale of the
maser distribution also strengthen the association between VLA--3 and the large-scale dynamics of the molecular envelope,
as traced by both jet ``footprints'' (HH objects and H$_2$ knots) and the wide-angle outflow seen at the sub-parsec
scale ($\rm \sim0.5~pc$ from VLA--3).
The other two objects indicated by water maser emission in Figure~\ref{puzzle} (the NE and MW cluster) were not
detected at cm and mm wavelengths. We give an upper limit to their flux density at 1.3~cm of about 0.5~mJy ($3\sigma$).
On the one hand, by assuming this radio continuum would come from a photoionized stellar wind, we can put an upper limit
on the spectral type of the exciting YSOs. Following equation~(1) of \citet{Rodriguez1983} which gives the rate
of ionizing photons required to fully ionize a constant velocity wind,  we get a value of $\rm N_L\lesssim 10^{47}~s^{-1}$
that would imply a ZAMS star with a spectral type later than B0.
On the other hand, the high, isotropic, maser-luminosity of the NE and MW objects ($\rm 2.5\times10^{-5}~L_{\odot}$ and
$\rm 3.3\times10^{-5}~L_{\odot}$, respectively) would suggest they are not even low-mass YSOs but possibly
late B-type stars, by comparison with the SE cluster luminosity ($\rm 1.2\times10^{-5}~L_{\odot}$) and the H$_2$O
luminosity usually associated with low-mass stars ($< 10^{-7}~L_{\odot}$; e.g., \citealt{Furuya2003} and their
erratum)\footnote{Note that from \citet{Trinidad2003},
the isotropic maser-luminosity associated with the VLA--2 radio source is of about $\rm 4\times10^{-5}~L_{\odot}$
(their Table~2) whereas VLA--1 does not show water maser emission.}.
In the formulation of \citet{Reid1988}, the maser luminosity is a consequence of the
dissipation in the ambient cloud of the mechanical energy of the outflow from a young star, and thus could be related to
the mass of the driving source.

\section{Summary and Conclusions}

\begin{figure*}
\centering
\includegraphics[angle=0,scale=0.9]{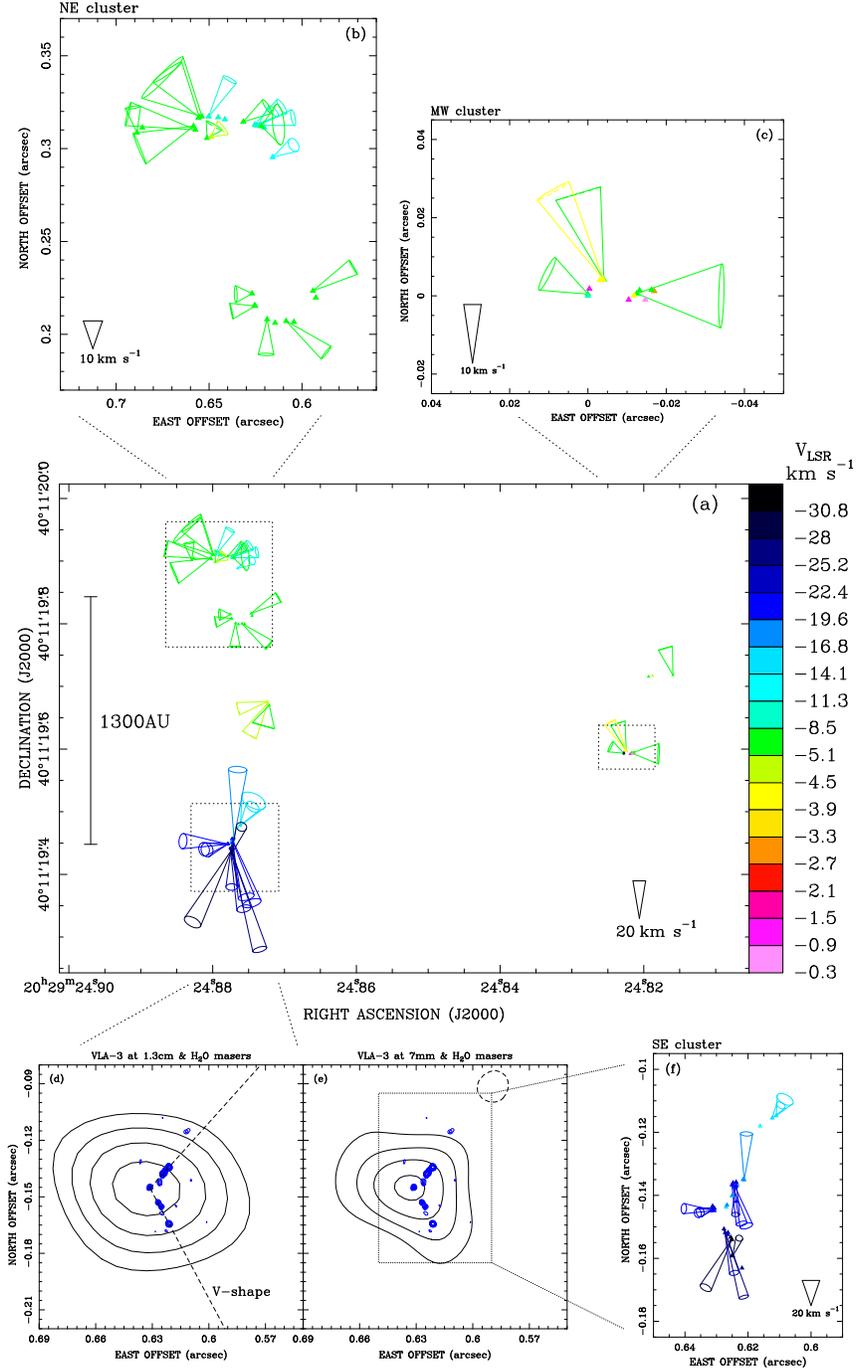}
\caption{{\tiny 22.2~GHz H$_2$O maser kinematics and radio continuum toward the VLA--3 source in AFGL~2591. \textit{Central Panel:} Absolute
positions (triangles) and internal proper motions of the 22~GHz H$_2$O maser features (see Table~\ref{distrib}). Colored cones are used to show both
the direction and the uncertainty (cone aperture) of the 3--D velocities of maser features. The proper-motion amplitude scale and linear-size scale
are given ($\rm 10~mas = 33~AU$). Different colors mark the maser LSR velocities according to the righthand side scale, centered on the systematic
velocity of the HMC (green equal to -5.7~km~s$^{-1}$). \textit{Upper Panels:} Details of the proper motions in the NE and MW water maser clusters.
Relative positions refer to the phase-reference feature~\#~4 in Table~\ref{distrib}. \textit{Bottom Panels:} from left to right, detail of the
extended water maser emission (blue contours) overlapped to the VLA--3 radio component of \citet{Trinidad2003}, in the same reference frame of the
upper panels. Blue contours are spaced by 3$\sigma$ starting from a 3$\sigma$ rms of 0.39~Jy~beam$^{-1}$. For the VLA--A radio continuum maps (black contours), plotted levels at 1.3~cm are spaced by 1$\sigma$ starting from a 3$\sigma$ rms, whereas for the 7~mm map contours are -3$\sigma$, 3$\sigma$, 5$\sigma$, 7$\sigma$, and 9$\sigma$ (see Table~\ref{cont}). A blow-up of the water
maser proper motions associated with VLA--3 is shown.}}
\label{puzzle}
\end{figure*}

\begin{figure*}
\centering
\includegraphics [angle=-90,scale=0.6]{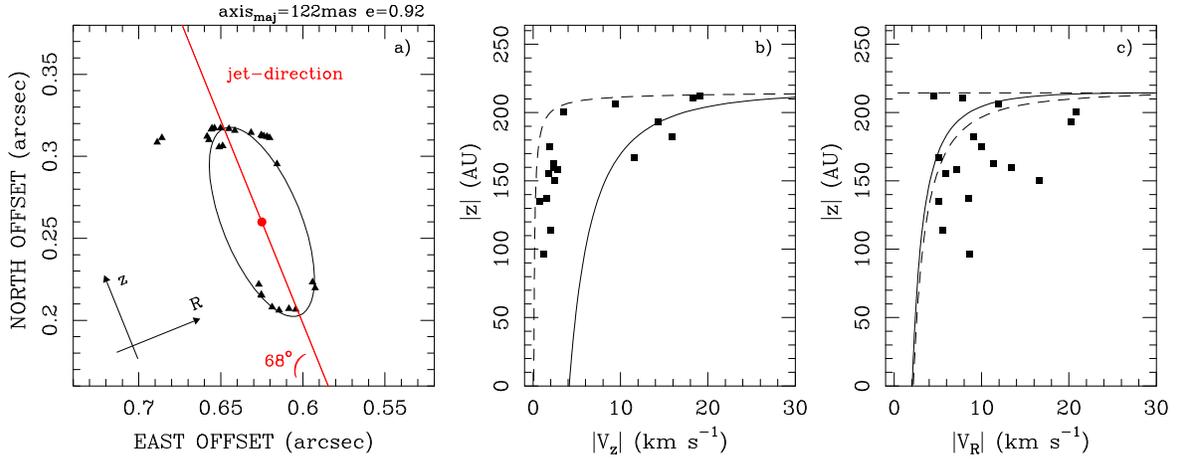}
\caption{Symmetry and velocity profiles of the NE water maser cluster. \textit{Left Panel:} Spatial distribution of
water maser features (triangles) superposed to the elliptical toy model used to define the new z--R reference frame. The major axis and
eccentricity of the ellipse are reported on top of the panel. \textit{Middle Panel:} Velocity profile of the water maser proper
motions along the z-direction in astronomical units. The lines represent the upper (solid line) and lower (dashed line) limits
to the bow shock model of \citet{Ostriker2001} discussed in Section~\ref{NEcluster}. \textit{Right Panel:} Velocity profile of the
transverse component, with respect the the z-direction, of the water maser proper motions as a function of z. The lines have the same
meaning as in the Middle Panel.}
\label{pv}
\end{figure*}

\begin{figure*}
\centering
\includegraphics [angle=0,scale=0.7]{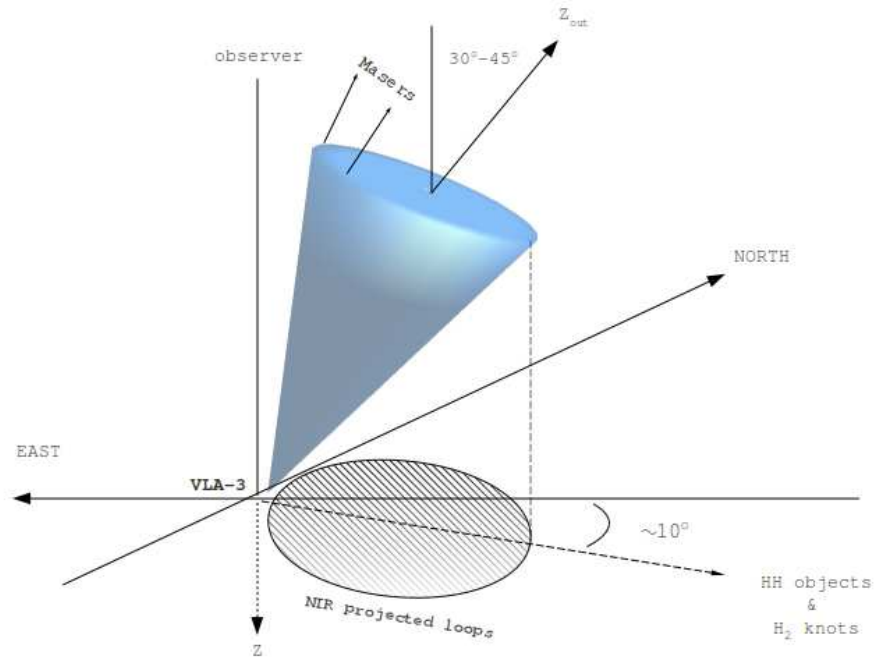}
\caption{Sketch of the blueshifted jet/outflow system in \AFGL\ emanating from the radio source VLA--3 and associated
to the SE water maser cluster (Figure~\ref{puzzle}d--f).
The inclination of the outflow axis about the line-of-sight ($\sim 30-45\degr$) and on the plane of the sky
($\sim 10\degr$) is reported. The NIR loops are shown as a cross section of the outflow projected along the
plane of the sky. The water maser emission associated with VLA--3 emerges from the outflow walls and mainly
expands toward the observer (Section~\ref{SEcluster}).}
\label{outflow}
\end{figure*}

We have presented multi-epoch, VLBA, H$_2$O maser observations toward the most compact radio source (i.e. VLA--3)
in the high-mass star-forming region \AFGL. We have also compared the maser distribution with the brightness
structure of the radio continuum at 1.3 and 0.7~cm observed with the VLA. Our main conclusions can be
summarized as follows:

\begin{enumerate}

\item We have detected three main clusters of water maser emission above a detection limit of 0.05~Jy~beam$^{-1}$
($5\sigma$) within $\rm \sim 2000~AU$ on the sky from the VLA--3 radio source (the SE water maser cluster).
Each water maser cluster shows internal expansion with velocities of a few tens of \kms\ (Figure~\ref{puzzle}a).
A global view of the star formation activity in \AFGL\ shows at least two spatial scales of star formation,
one projected across 0.1~pc on the sky and another one at a ten times smaller scale as indicated by
water maser emission.

\item The kinematics of the NE water maser cluster shows characteristics of two symmetric, jet-driven,
bow shocks expanding about 200~AU from a still undetected protostar, with a velocity of about 66~\kms\
(Figure~\ref{puzzle}b). Our analysis of the water maser proper motions, compared to different outflow
models, suggests that the primary wind may be a combination of a jet and a lower-velocity wind tilted
about $20 \degr$ from the jet axis. The short dynamical age (14~yr) of these expanding motions suggests
they are probably related to recurrent ejection events from the central protostar(s).

\item Our analysis of the radio continuum emission compared with the IR luminosity of the region shows that
VLA--3 is likely a ZAMS star with a spectral type between O9--O6 and a mass in the range $\rm 20-38~M_{\odot}$.
This young stellar object dominates the IR luminosity of the region. The water maser emission associated with VLA--3 expands toward
the observer with velocities up to 50~\kms\ and traces a V--shape along the plane of the sky (Figure~\ref{puzzle}d--f).
The orientation of the maser V--shape is aligned with the high-velocity wings of the CO outflow and the
Herbig-Haro objects at a few tenth of pc from the star. The wide opening angle of the maser V--shape ($\sim 110\degr$)
agrees with that of the outflow as inferred from NIR emission along the edges of the outflow cavity.
This water maser emission is likely amplified along the walls of the blueshifted outflow lobe from VLA--3 (Figure~\ref{outflow}).

\end{enumerate}

\acknowledgments
We thank J. M. Torrelles and A. P. Lobanov for helpful discussions in preparation.
This work was partially funded by the ERC Advanced Investigator Grant GLOSTAR (247078).
KLJR is funded by an ASI fellowship under contract number I/005/07/1.

{\it Facilities:} \facility{VLBA}.

\clearpage

%\begin{landscape}
%\end{landscape}

\clearpage

\clearpage

\clearpage

\begin{deluxetable}{lclcrccr}
\tabletypesize{\scriptsize}
%%\rotate
\tablecaption{Radio continuum toward the VLA--3 component in AFLG~2591 \label{cont}}
\tablewidth{0pt}
\tablehead{
\colhead{ } &\colhead{ } &\colhead{ } &\colhead{ } &   \multicolumn{2}{c}{Peak position}  &\colhead{ } &\colhead{ } \\
\colhead{Telescope} &\colhead{$\lambda$} & \colhead{HPBW} & Image rms &  \colhead{$\Delta \rm x$} & \colhead{$\Delta \rm y$} &
\colhead{F$_{\rm peak}$} & \colhead{Deconvolution} \\
\colhead{ }  & \colhead{(cm)}  & \colhead{$ \rm mas \times mas \ at \ \degr$} & \colhead{(mJy beam$^{-1}$)} & \colhead{(mas)} &
\colhead{(mas)} & \colhead{(mJy beam$^{-1}$)} & \colhead{$ \rm mas \times mas \ at \ \degr$} \\}

\startdata

VLA--A  & 1.3 &  $ 80 \times 80 $              & 0.19 & $633 \pm 10$ & $-145 \pm 10$  & 1.24 &  $ 88\times38 \ at \   88\degr$  \\
VLA--A  & 0.7 &  $ 45 \times 38 \ at -34\degr$ & 0.11 & $633 \pm 10$ & $-145 \pm 10$  & 0.95 & 65~mas for slice at   75$\degr$  \\
\nodata & \nodata & \nodata & \nodata & \nodata &  \nodata & \nodata &        48~mas for slice at  $-48\degr$  \\
\nodata & \nodata & \nodata & \nodata & \nodata &  \nodata & \nodata &        62~mas for slice at $-159\degr$  \\

\enddata
%% Text for table notes should follow after the \enddata but before
%% the \end{deluxetable}. Make sure there is at least one \tablenotemark{a}
%% in the table for each \tablenotetext.
%%
\tablecomments{The 1.3 and 0.7~cm VLA--A measurements are from archival data \citep{Trinidad2003,vanderTak2005}.
The peak position in columns 5--6 is the relative position with respect to the reference feature of Table~\ref{distrib}. Column 8 gives the
deconvolved FWHM sizes of the Gaussian fits to the 1.3~cm emission and along the three axes of the 0.7~cm emission (position angles
east of north).}
%%\tablenotetext{a}{ ... }
%%\tablenotetext{b}{ ... }

\end{deluxetable}

\begin{deluxetable}{lcrrrrrr}
\tabletypesize{\scriptsize}
%\rotate
\tablecaption{Parameters of VLBA 22.2~GHz water maser features.\label{distrib}}
\tablewidth{0pt}
\tablehead{
\colhead{Feature} & \colhead{Detection} &\colhead{V$_{\rm LSR}$} & \colhead{F$_{\rm peak}$} &  \colhead{$\Delta \rm x$} &
\colhead{$\Delta \rm y$} & \colhead{\rm V$_{\rm x}$}    & \colhead{V$_{\rm y}$}  \\
\colhead{\#} & \colhead{(epochs)} &\colhead{(\kms)}    & \colhead{(Jy beam$^{-1}$)}    &  \colhead{(mas)} & \colhead{(mas)} &
\colhead{(\kms)} & \colhead{(\kms)} \\}

\startdata

   \multicolumn{8}{c}{\textbf{North West Cluster} } \\
1  &  1                  & $-5.00$ & 14.87 &  $ -39.72 \pm 0.11 $ & $ 122.69 \pm 0.05 $  &   \nodata   &  \nodata    \\
2  & 1,2,3,\underline{4} & $-5.42$ & 14.17 &  $ -78.51 \pm 0.06 $ & $ 125.13 \pm 0.05 $  & $ 4.0 \pm 2.9 $ & $ 14.2 \pm 2.3 $ \\ % 16\%
3  &  1                  & $-5.00$ &  0.55 &  $ -46.13 \pm 0.08 $ & $ 124.11 \pm 0.05 $  &   \nodata   &  \nodata    \\
   \multicolumn{8}{c}{\textbf{Middle West Cluster} } \\
4  & 1,2,\underline{3},4 &$-11.74$ & 20.60 &  $   0   \pm 0.06 $ & $  0   \pm 0.09 $ &   \nodata   &  \nodata    \\
5  & \underline{1},2,3,4 & $-5.42$ & 13.41 &  $ -4.07 \pm 0.12 $ & $ 4.16 \pm 0.06 $ & $ 4.3 \pm 3.2 $ & $ 14.6 \pm 2.1 $ \\ % 15\%
6  & 1,\underline{2},3,4 & $-6.68$ &  8.29 &  $  0.02 \pm 0.05 $ & $ 0.47 \pm 0.07 $ & $ 6.9 \pm 1.9 $ & $  3.4 \pm 2.2 $ \\ % 26\%
7  & 1,2,\underline{3}   & $-7.10$ &  2.98 &  $-12.47 \pm 0.07 $ & $ 0.59 \pm 0.06 $ & $-14.4 \pm 3.3 $ & $ -0.3 \pm 4.1 $ \\ % 23\%
8  & 1,2,3,\underline{4} & $-4.57$ &  1.60 &  $ -3.91 \pm 0.07 $ & $ 4.18 \pm 0.07 $ & $ 8.5 \pm 2.1 $ & $ 15.0 \pm 2.2 $ \\ % 13\%
9  & 1                   & $-1.20$ &  0.24 &  $ -0.37 \pm 0.06 $ & $ 1.79 \pm 0.08 $ &   \nodata   &  \nodata    \\
10 &   4                 & $-2.89$ &  0.24 &  $-16.96 \pm 0.08 $ & $ 1.19 \pm 0.12 $ &   \nodata   &  \nodata    \\
11 & 1                   & $-4.15$ &  0.18 &  $-11.78 \pm 0.05 $ & $ 0.10 \pm 0.06 $ &   \nodata   &  \nodata    \\
12 & 1                   & $-4.58$ &  0.17 &  $ -3.04 \pm 0.05 $ & $ 3.99 \pm 0.06 $ &   \nodata   &  \nodata    \\
13 &   4                 & $-0.36$ &  0.14 &  $-14.65 \pm 0.07 $ & $-0.97 \pm 0.07 $ &   \nodata   &  \nodata    \\
14 &   4                 & $-6.26$ &  0.09 &  $-16.26 \pm 0.07 $ & $ 1.51 \pm 0.10 $ &   \nodata   &  \nodata    \\
15 & 1                   & $-7.94$ &  0.09 &  $-13.20 \pm 0.06 $ & $ 1.38 \pm 0.08 $ &   \nodata   &  \nodata    \\
16 & 1                   & $-1.20$ &  0.06 &  $-10.40 \pm 0.07 $ & $-1.00 \pm 0.06 $ &   \nodata   &  \nodata    \\
   \multicolumn{8}{c}{\textbf{North East Cluster} } \\
17 & 1,2,3,\underline{4} & $-5.00$ & 22.25 &  $ 651.05 \pm 0.06 $ & $ 305.69 \pm 0.06 $  & $ -2.5 \pm 1.8 $ & $  4.4 \pm 2.1 $\\ % 41\%
18 & 1,2,\underline{3},4 & $-9.21$ &  6.01 &  $ 650.18 \pm 0.05 $ & $ 317.22 \pm 0.06 $  & $ -7.5 \pm 1.7 $ & $ 13.2 \pm 2.0 $\\ % 13\%
19 & \underline{1},2,3,4 &$-12.58$ &  5.44 &  $ 615.75 \pm 0.06 $ & $ 295.33 \pm 0.06 $  & $ -7.5 \pm 2.0 $ & $  4.3 \pm 2.5 $\\ % 25\%
20 & \underline{1},2,3,4 & $-7.53$ &  4.52 &  $ 618.75 \pm 0.10 $ & $ 208.02 \pm 0.10 $  & $  0.3 \pm 2.2 $ & $-12.5 \pm 2.4 $\\ % 19\%
21 & 1,2,3,\underline{4} & $-8.79$ &  2.89 &  $ 624.70 \pm 0.06 $ & $ 312.39 \pm 0.09 $  & $-13.2 \pm 2.1 $ & $  2.8 \pm 2.1 $\\ % 16\%
22 & 1,2,\underline{3},4 & $-8.79$ &  2.29 &  $ 623.68 \pm 0.11 $ & $ 312.34 \pm 0.05 $  & $ -5.5 \pm 1.7 $ & $  5.2 \pm 2.0 $\\ % 25\%
23 & 4                   & $-9.21$ &  2.03 &  $ 645.05 \pm 0.06 $ & $ 316.82 \pm 0.06 $  &   \nodata   &  \nodata    \\
24 & 1,2,\underline{3}   & $-8.37$ &  2.01 &  $ 655.31 \pm 0.10 $ & $ 317.05 \pm 0.07 $  & $ 11.3 \pm 4.2 $ & $16.0 \pm 4.3 $\\ % 22\%
25 & \underline{2},3     & $-8.37$ &  1.85 &  $ 654.88 \pm 0.08 $ & $ 317.44 \pm 0.06 $  &   \nodata   &  \nodata    \\
26 & \underline{1},2,3,4 & $-7.53$ &  1.60 &  $ 608.52 \pm 0.06 $ & $ 207.16 \pm 0.06 $  & $-14.3 \pm 1.8 $ & $-11.3 \pm 2.2 $\\ % 11\%
27\tablenotemark{a} & \underline{1},2,3,4 & $-8.37$ &  1.35 &  $ 620.27 \pm 0.19 $ & $ 311.32 \pm 0.16 $  &   \nodata   &  \nodata    \\
28 & 1,2,3,\underline{4} & $-7.11$ &  0.69 &  $ 625.15 \pm 0.05 $ & $ 215.20 \pm 0.05 $  & $  8.5 \pm 1.8 $ & $ -1.7 \pm 2.1 $\\ % 21\%
29\tablenotemark{a} & 1,2,\underline{3}   & $-6.69$ &  0.53 &  $ 625.44 \pm 0.05 $ & $ 215.62 \pm 0.06 $  &   \nodata   &  \nodata    \\
30 & 1,\underline{2},3,4 & $-5.00$ &  0.52 &  $ 648.84 \pm 0.08 $ & $ 306.30 \pm 0.06 $  & $ -4.0 \pm 2.0 $ & $ 3.1 \pm 2.0 $\\ % 40\%
31\tablenotemark{a} & \underline{1},2,3,4 & $-9.21$ &  0.47 &  $ 641.48 \pm 0.09 $ & $ 315.73 \pm 0.05 $  &   \nodata   &  \nodata    \\
32 & 1,2,3,\underline{4} & $-8.27$ &  0.42 &  $ 631.55 \pm 0.05 $ & $ 314.48 \pm 0.06 $  & $ -8.5 \pm 1.8 $ & $ 5.4 \pm 2.2 $\\ % 20\%
33 & \underline{1},2,3,4 & $-9.63$ &  0.40 &  $ 625.42 \pm 0.09 $ & $ 312.92 \pm 0.10 $  & $ -9.7 \pm 2.1 $ & $ 6.4 \pm 2.2 $\\ % 18\%
34 & \underline{1},2,3   & $-8.37$ &  0.39 &  $ 653.85 \pm 0.06 $ & $ 317.29 \pm 0.24 $  & $ 14.1 \pm 3.3 $ & $14.0 \pm 8.7 $\\ % 33\%
35 & \underline{1},2,3,4 & $-7.53$ &  0.39 &  $ 594.08 \pm 0.05 $ & $ 223.24 \pm 0.13 $  & $-14.4 \pm 1.8 $ & $ 8.5 \pm 2.2 $\\ % 12\%
36 & 2,3,\underline{4}   & $-8.37$ &  0.31 &  $ 657.75 \pm 0.06 $ & $ 309.98 \pm 0.06 $  & $ 24.1 \pm 3.0 $ & $ 5.7 \pm 3.2 $\\ % 12\%
37 & 1,\underline{2},3   & $-6.26$ &  0.24 &  $ 658.35 \pm 0.07 $ & $ 312.14 \pm 0.09 $  & $ 18.0 \pm 3.3 $ & $-11.0 \pm 4.7 $\\ % 17\%
38 & \underline{1},2,3,4 & $-6.69$ &  0.23 &  $ 688.86 \pm 0.08 $ & $ 308.38 \pm 0.14 $  & $  1.4 \pm 1.8 $ & $ 9.0 \pm 2.2 $\\ % 24\%
39\tablenotemark{a} & \underline{1},2,3   & $-6.69$ &  0.22 &  $ 685.88 \pm 0.05 $ & $ 311.26 \pm 0.05 $  &   \nodata   &  \nodata    \\
40 & 1,2,\underline{3},4 & $-6.69$ &  0.19 &  $ 626.85 \pm 0.05 $ & $ 221.95 \pm 0.06 $  & $ 5.8 \pm 1.8 $ & $ -0.2 \pm 2.1 $\\ % 31\%
41 & 2,3,\underline{4}   & $-7.53$ &  0.19 &  $ 620.88 \pm 0.25 $ & $ 311.72 \pm 0.08 $  & $-6.1 \pm 3.7 $ & $  0.5 \pm 3.0 $\\ % 60\%
42 & 1                   & $-7.95$ &  0.16 &  $ 614.48 \pm 0.05 $ & $ 206.15 \pm 0.05 $  &   \nodata   &  \nodata    \\
43 & 4                   & $-7.95$ &  0.13 &  $ 592.54 \pm 0.08 $ & $ 219.73 \pm 0.14 $  &   \nodata   &  \nodata    \\
44 & 4                   & $-7.53$ &  0.10 &  $ 604.34 \pm 0.08 $ & $ 206.63 \pm 0.07 $  &   \nodata   &  \nodata    \\
   \multicolumn{8}{c}{\textbf{Middle East Cluster} } \\
45 & \underline{1},2,3,4 & $-4.57$ & 3.14 &  $ 567.48 \pm 0.06 $ & $ 80.27 \pm 0.10 $  & $ 9.7 \pm 2.2 $ & $ -16.9 \pm 3.5 $\\ % 17\%
46 & \underline{1},2,3,4 & $-5.00$ & 0.58 &  $ 568.34 \pm 0.06 $ & $ 82.61 \pm 0.11 $  & $15.5 \pm 1.9 $ & $  -3.8 \pm 2.7 $\\ % 13\%
47 & \underline{1},2,3   & $-5.84$ & 0.28 &  $ 565.23 \pm 0.06 $ & $ 76.66 \pm 0.06 $  & $ 3.0 \pm 3.3 $ & $ -10.8 \pm 4.2 $\\ % 37\%
   \multicolumn{8}{c}{\textbf{South East Cluster} } \\
48 & 2,\underline{3},4 & $-24.38$ & 4.83 &  $ 626.73 \pm 0.05 $ & $ -153.13 \pm 0.05 $  & $-7.2 \pm 3.2 $ & $ -31.9 \pm 3.9 $\\ % 12\%
49 &  2,\underline{3}  & $-27.75$ & 4.37 &  $ 621.15 \pm 0.07 $ & $ -164.35 \pm 0.06 $  &   \nodata   &  \nodata    \\
50 & 1                 & $-16.38$ & 4.32 &  $ 623.54 \pm 0.12 $ & $ -136.49 \pm 0.07 $  &   \nodata   &  \nodata    \\
51 & 2,\underline{3},4 & $-19.32$ & 2.05 &  $ 624.58 \pm 0.08 $ & $ -137.92 \pm 0.11 $  & $ -7.7 \pm 3.4 $ & $ -30.7 \pm 3.8 $\\ % 12\%
52 & \underline{2},3,4 & $-19.75$ & 1.71 &  $ 620.92 \pm 0.15 $ & $ -134.25 \pm 0.11 $  & $ -1.9 \pm 4.4 $ & $  36.4 \pm 3.8 $\\ % 11\%
53 &  \underline{2},3  & $-18.90$ & 1.59 &  $ 621.43 \pm 0.07 $ & $ -134.56 \pm 0.06 $  &   \nodata   &  \nodata    \\
54 & 2,3,\underline{4} & $-25.65$ & 1.52 &  $ 626.89 \pm 0.06 $ & $ -152.60 \pm 0.09 $  & $-15.7 \pm 3.1 $ & $ -54.0 \pm 3.3 $\\ % 6\%
55 & \underline{2},3     & $-17.22$ & 1.23 &  $ 623.17 \pm 0.06 $ & $-135.73 \pm 0.06 $  &   \nodata   &  \nodata    \\
56\tablenotemark{a} & 2,3,\underline{4} & $-18.06$ & 0.97 &  $ 626.23 \pm 0.08 $ & $ -143.15 \pm 0.09 $  &   \nodata   &  \nodata    \\
57 & \underline{2},3,4   & $-20.59$ & 0.94 &  $ 624.09 \pm 0.10 $ & $-137.34 \pm 0.10 $  & $ -10.5 \pm 3.9 $ & $ -34.2 \pm 4.4 $\\ % 12\%
58 & \underline{1},2,3,4 & $-14.27$ & 0.82 &  $ 612.29 \pm 0.23 $ & $-115.46 \pm 0.10 $  & $  -6.8 \pm 3.6 $ & $  10.1 \pm 2.9 $\\ % 26\%
59 &   2,3,\underline{4} & $-29.44$ & 0.79 &  $ 624.39 \pm 0.06 $ & $ -159.05\pm 0.07 $  & $  -5.5 \pm 3.3 $ & $  14.0 \pm 3.6 $\\ % 24\%
60 & 1                 & $-21.01$ & 0.70 &  $ 631.75 \pm 0.14 $ & $ -144.95 \pm 0.05 $  &   \nodata   &  \nodata    \\
61 & \underline{1},2,3 & $-30.70$ & 0.61 &  $ 625.12 \pm 0.06 $ & $ -153.96 \pm 0.07 $  & $ 20.7 \pm 3.6 $ & $ -38.4 \pm 4.4 $\\ % 10\%
62 & 1                 & $-20.17$ & 0.50 &  $ 623.75 \pm 0.07 $ & $ -137.19 \pm 0.08 $  &   \nodata   &  \nodata    \\
63 & 2,\underline{3},4 & $-21.01$ & 0.45 &  $ 631.22 \pm 0.13 $ & $ -145.12 \pm 0.05 $  & $ 24.0 \pm 5.2 $ & $  1.6 \pm 3.4 $\\ % 22\%
64 & 2,3,\underline{4} & $-23.96$ & 0.41 &  $ 623.51 \pm 0.06 $ & $ -136.87 \pm 0.07 $  & $  0.4 \pm 3.2 $ & $-25.7 \pm 3.3 $\\ % 13\%
65 & 1                 & $-29.44$ & 0.38 &  $ 625.51 \pm 0.06 $ & $ -153.29 \pm 0.06 $  &   \nodata   &  \nodata    \\
66 &  2,\underline{3}  & $-15.53$ & 0.36 &  $ 626.12 \pm 0.05 $ & $ -141.52 \pm 0.10 $  &   \nodata   &  \nodata    \\
67 & 1                 & $-22.28$ & 0.31 &  $ 631.30 \pm 0.07 $ & $ -144.40 \pm 0.06 $  &   \nodata   &  \nodata    \\
68 & \underline{1},2,3 & $-14.69$ & 0.24 &  $ 611.01 \pm 0.13 $ & $ -114.73 \pm 0.12 $  & $  -6.9 \pm 4.9 $ & $ 13.0 \pm 5.4 $\\ % 36\%
69 & \underline{2},3,4 & $-21.85$ & 0.21 &  $ 630.94 \pm 0.05 $ & $ -144.92 \pm 0.05 $  & $  12.7 \pm 4.0 $ & $ -2.7 \pm 3.5 $\\ % 31\%
70 & 1                 & $-22.70$ & 0.20 &  $ 631.24 \pm 0.05 $ & $ -143.37 \pm 0.06 $  &   \nodata   &  \nodata    \\
71 & 1                 & $-15.53$ & 0.18 &  $ 626.92 \pm 0.06 $ & $ -143.72 \pm 0.07 $  &   \nodata   &  \nodata    \\
72 & 2,\underline{3},4 & $-22.27$ & 0.18 &  $ 630.73 \pm 0.08 $ & $ -144.46 \pm 0.10 $  & $ 11.2 \pm 3.8 $ & $-3.2 \pm 4.3 $\\ % 33\%
73 & 4                 & $-21.01$ & 0.16 &  $ 623.82 \pm 0.08 $ & $ -136.98 \pm 0.09 $  &   \nodata   &  \nodata    \\
74 & 4                 & $-17.22$ & 0.14 &  $ 624.97 \pm 0.13 $ & $ -140.05 \pm 0.09 $  &   \nodata   &  \nodata    \\
75 & 4                 & $-23.96$ & 0.10 &  $ 626.35 \pm 0.06 $ & $ -151.82 \pm 0.07 $  &   \nodata   &  \nodata    \\
76 & 4                 & $-25.22$ & 0.09 &  $ 626.43 \pm 0.07 $ & $ -152.47 \pm 0.10 $  &   \nodata   &  \nodata    \\
77 &  2,\underline{3}  & $-33.65$ & 0.09 &  $ 624.83 \pm 0.06 $ & $ -155.15 \pm 0.07 $  &   \nodata   &  \nodata    \\
78 & 1                 & $-13.01$ & 0.07 &  $ 616.11 \pm 0.07 $ & $ -118.00 \pm 0.09 $  &   \nodata   &  \nodata    \\
79 & 4                 & $-22.28$ & 0.06 &  $ 623.72 \pm 0.06 $ & $ -141.92 \pm 0.10 $  &   \nodata   &  \nodata    \\
80 & 1                 & $-26.91$ & 0.05 &  $ 625.20 \pm 0.20 $ & $ -159.19 \pm 0.65 $  &   \nodata   &  \nodata    \\
 & & & & & & \\
\multicolumn{8}{c}{\textbf{Reference Feature \#~4: absolute position on 2009 May 6} } \\
\hline\hline
 & & & & & & \\
& & \multicolumn{2}{c}{R.A.~(J2000)}        &  \multicolumn{2}{c}{Decl.~(J2000)}  & & \\
& & \multicolumn{2}{c}{$20^h29^m24\fs8228$} &  \multicolumn{2}{c}{$40\degr11'19\farcs593$} & & \\

\enddata

\tablecomments{ For each identified feature belonging to a given cluster of maser emission, the label (given in Column~1) increases with decreasing
brightness and the different epochs of detection are reported in Column~2. Columns~3 and~4 report the LSR velocity and brightness of the brightest
spot of each feature, observed at the epoch underlined in Column~2. Columns~5 and~6 give the position offset at the first epoch of detection
relative to the feature~\#~4 (centered on the phase-reference maser channel \#~145) in the east and north directions, respectively.
The uncertainties give the intensity-weighted standard deviation of the spots distribution within a feature and was combined in quadrature with a
conservative 50~$\mu$as uncertainty. This 50~$\mu$as plateau accounts for deviations from a Gaussian fit assumption.
The absolute position of the reference maser feature is reported at the bottom of the table and is accurate to within $\pm 1$~mas.
Columns~7 and~8 report the projected components of the feature proper motion relative to the feature~\#~4, along the east and north
directions, respectively. A proper motion of 1~mas~yr$^{-1}$ corresponds to 15.6~km~s$^{-1}$ at a distance of 3.3~kpc.}
\tablenotetext{a}{ Features with too uncertain proper motions ($> 70\%$)}.

\end{deluxetable}

\end{document}